\documentclass[conference]{IEEEtran}

\usepackage{cite}
\usepackage{graphicx}
\usepackage{subcaption}
\usepackage{xcolor}
\usepackage{pgfplots}
\usepackage{multirow}
\usepackage{upgreek}
\usepackage{amssymb}
\usepackage{amsmath}
\usepackage{amsthm}
\usepackage{cases}
\usepackage{pifont}
\usepackage{bm}
\usepackage{bbm}

\usepackage{tabularx}
\usepackage{booktabs}
\usepackage{array}
\usepackage{float}
\usepackage{algorithmic}
\usepackage[linesnumbered,ruled,vlined]{algorithm2e}
\usepackage{changes}
\usepackage{hhline}
\usepackage{soul}
\usetikzlibrary{arrows,positioning,shapes.geometric,spy}

\newcommand{\fixme}[2]{\ifx&#2&{\leavevmode\color{red}#1}\else{\leavevmode\color{red}FIXME\{}#1{\leavevmode\color{red}\}}\footnote{{\leavevmode\color{red}#2}}\PackageWarning{Fixme}{#1: #2}\fi}

\SetCommentSty{mycommfont}

\DeclareMathOperator*{\argmin}{arg\,min}
\DeclareMathOperator*{\argmax}{arg\,max}
\DeclareMathOperator*{\sgn}{sgn}
\DeclareMathOperator{\PM}{PM}

\begin{document}

\title{
Deep-Learning-Aided Successive-Cancellation Decoding of Polar Codes
}

\author{
\IEEEauthorblockN{Seyyed Ali Hashemi\IEEEauthorrefmark{1}, Nghia Doan\IEEEauthorrefmark{2}, Thibaud Tonnellier\IEEEauthorrefmark{2}, Warren J. Gross\IEEEauthorrefmark{2}}
\IEEEauthorblockA{\IEEEauthorrefmark{1}Department of Electrical Engineering, Stanford University, USA}
\IEEEauthorblockA{\IEEEauthorrefmark{2}Department of Electrical and Computer Engineering, McGill University, Canada}
\IEEEauthorblockA{ahashemi@stanford.edu, nghia.doan@mail.mcgill.ca, thibaud.tonnellier@mcgill.ca, warren.gross@mcgill.ca}
}

\maketitle
\begin{abstract}
A deep-learning-aided successive-cancellation list (DL-SCL) decoding algorithm for polar codes is introduced with deep-learning-aided successive-cancellation (DL-SC) decoding being a specific case of it. The DL-SCL decoder works by allowing additional rounds of SCL decoding when the first SCL decoding attempt fails, using a novel bit-flipping metric. The proposed bit-flipping metric exploits the inherent relations between the information bits in polar codes that are represented by a correlation matrix. The correlation matrix is then optimized using emerging deep-learning techniques. Performance results on a polar code of length $128$ with $64$ information bits concatenated with a $24$-bit cyclic redundancy check show that the proposed bit-flipping metric in the proposed DL-SCL decoder requires up to $66\%$ fewer multiplications and up to $36\%$ fewer additions, without any need to perform transcendental functions, and by providing almost the same error-correction performance in comparison with the state of the art.
\end{abstract}

\IEEEpeerreviewmaketitle
\section{Introduction} \label{sec:intro}

Polar codes represent a class of error-correcting codes that are proven to achieve channel capacity for any binary symmetric channel under the low-complexity successive-cancellation (SC) decoding \cite{arikan}. Recently, polar codes are selected for use in the enhanced mobile broadband (eMBB) control channel of the fifth generation of cellular technology (5G standard), where codes with short block length are used \cite{3gpp_report}. The error-correction performance of short polar codes under SC decoding does not satisfy the requirements of the 5G standard. SC list (SCL) decoding was introduced in \cite{tal_list} to improve the error-correction performance of SC decoding by keeping a list of candidate message words at each decoding step. In addition, it was observed that under SCL decoding, the error-correction performance is significantly improved when the polar code is concatenated with a cyclic redundancy check (CRC) code \cite{tal_list}. However, the decoding complexity of SCL grows as the list size increases.

Unlike SCL decoding, SC flip (SCF) decoding \cite{SCF} performs multiple SC decoding attempts in series where in each attempt, the first-order erroneous information bit in the initial SC decoding attempt is flipped. Similar to SCL decoding, SCF decoding uses a CRC code to determine whether a decoding attempt is successful or not and a bit-flipping metric is used to identify the erroneous information bit. Several methods have been proposed to improve the error-correction performance of SCF \cite{Carlo_SCFlip,PSCF-ICC18,SCFlip_TCOM18}. However, the bit-flipping metric of a given information bit is oversimplified where only the log-likelihood ratio (LLR) corresponding to that bit is considered. To overcome this problem, dynamic SCF (DSCF) decoding \cite{DSCF} defines a more accurate bit-flipping metric, which utilizes the LLR values of all the previously decoded information bits. It was shown in \cite{DSCF} that at practical signal-to-noise ratio (SNR) values, DSCF decoding can achieve an error-correction performance comparable to SCL decoding, while maintaining an average decoding complexity close to that of SC decoding. But the bit-flipping metric in DSCF decoding requires costly exponential and logarithmic computations, which hinders the algorithm to be efficiently implemented in hardware.

In this paper, the likelihood of the correct decoding of each information bit under SC or SCL decoding is estimated by exploiting the inherent correlations among all the information bits. These correlations are expressed in the form of a trainable correlation matrix. Consequently, a bit-flipping metric based on the proposed correlation matrix is introduced. It only requires the computation of multiplication and addition operations in the LLR domain, preventing completely the needs to use costly transcendental functions as required by DSCF decoding. Motivated by recent developments that exploit deep learning (DL) to decode polar codes \cite{Cammerer_ScaleDL,Xu,Doan_SPAWC18,Doan_ICC19,Doan_SiPS19}, DL techniques are applied to optimize the correlation matrix. Thus the proposed decoding algorithm is called deep-learning-aided SCL (DL-SCL) decoding with DL-SC decoding being its specific case when the list size is one. Performance results on a polar code of length $128$ with $64$ information bits concatenated with a $24$-bit CRC show that the proposed bit-flipping metric in the proposed DL-SCL decoder requires up to $66\%$ fewer multiplications and up to $36\%$ fewer additions in comparison with the decoder that uses the bit-flipping metric in \cite{DSCF}. Moreover, the proposed decoder with the proposed bit-flipping metric does not require to perform any transcendental functions and can provide almost the same error-correction performance in comparison with the decoder that uses the bit-flipping metric in \cite{DSCF}.

\section{Preliminaries}

\label{sec:polar}

\subsection{Polar Codes, SC Decoding, and SCL Decoding}
\label{sec:polar:polar}

A polar code $\mathcal{P}(N,K)$ of block length $N$ with $K$ information bits is derived as $\bm{x} = \bm{u}\bm{G}^{\otimes n}$, where $\bm{x} = \{x_0,x_1,\ldots,x_{N-1}\}$ is the polar codeword, $\bm{u} = \{u_0,u_1,\ldots,u_{N-1}\}$ is the message word, $\bm{G}^{\otimes n}$ is the $n$-th Kronecker power of the polarizing matrix $\bm{G}=\bigl[\begin{smallmatrix} 1&0\\ 1&1 \end{smallmatrix} \bigr]$, and $n = \log_2 N$. The vector $\bm{u}$ consists of a set $\mathcal{A}$ of the indices of $K$ information bits and a set $\mathcal{A}^c$ of the indices of $N-K$ frozen bits. The positions of frozen bits are known to both the encoder and the decoder, and their values are set to $0$. In this paper, binary phase-shift keying (BPSK) modulation technique is considered. Therefore, the received signals of the transmitted codeword are represented as $\bm{y}=(\mathbf{1}-2\bm{x})+\bm{z}$, where $\mathbf{1}$ is an all-one vector of size $N$, and $\bm{z} \in \mathbb{R}^N$ is the additive white Gaussian noise (AWGN) vector with variance $\sigma^2$ and zero mean. The LLR vector of the received signal is then given as $\bm{L}_{n} = \frac{2\bm{y}}{\sigma^2}$.


\begin{figure}[t]
	\centering
	\begin{subfigure}[b]{0.5\textwidth}
		\centering
		\includegraphics[width=0.65\linewidth]{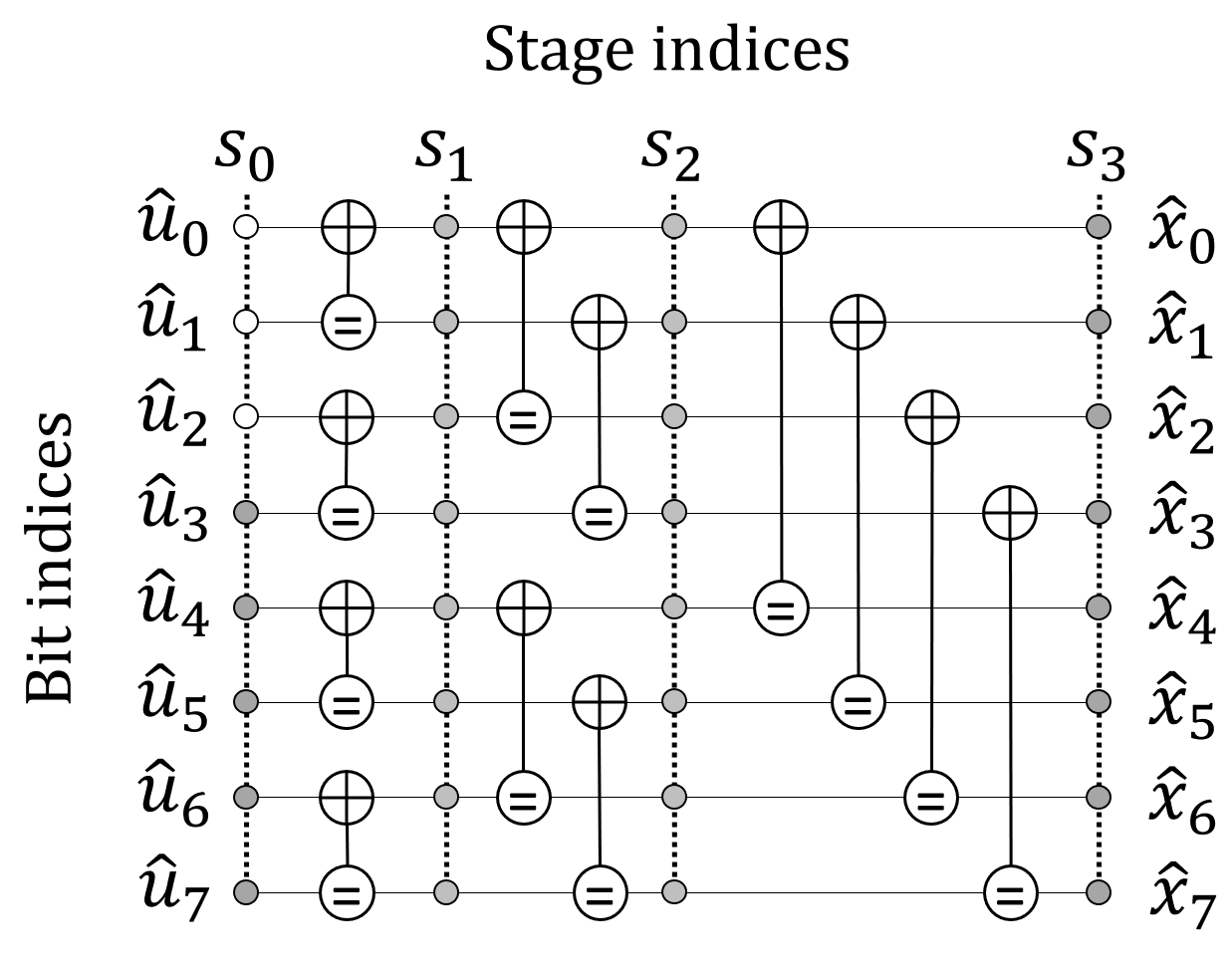}
		\caption{}
		\vspace*{5pt}
		\label{fig:SCGraph}
	\end{subfigure}
	\begin{subfigure}[b]{0.5\textwidth}			
		\centering
		\hspace*{15pt}
		\includegraphics[width=0.65\linewidth]{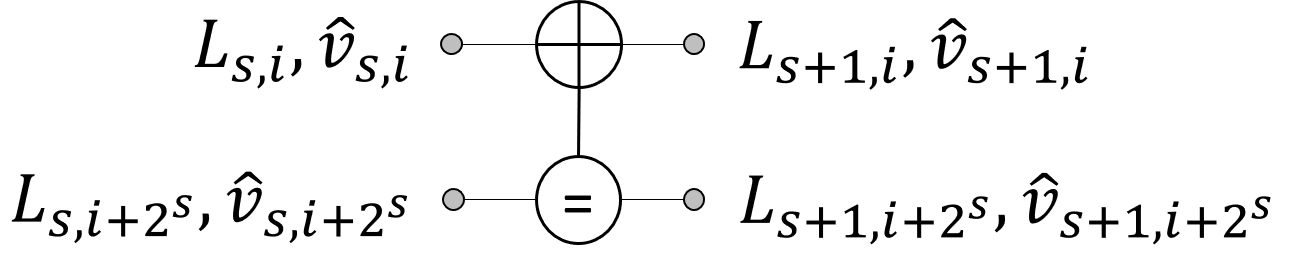}
		\caption{}
		\label{fig:SCPE}
	\end{subfigure}  
	\caption{(a) SC decoding on the factor graph of $\mathcal{P}(8,5)$ with $\mathcal{A}^c = \{0,1,2\}$, (b) a PE.}
\end{figure}

SC decoding can be illustrated on a polar code factor graph representation. Fig.~\ref{fig:SCGraph} shows an example of a factor graph for $\mathcal{P}(8,5)$. To obtain the estimated message word, the LLR values and the hard bit estimations are propagated through all the processing elements (PEs) in the factor graph that are depicted in Fig.~\ref{fig:SCPE}. A PE performs LLR computations as
\begin{align}
&L_{s,i} \!=\! \min(|L_{s+1,i}|,|L_{s+1,i+2^s}|)\sgn(L_{s+1,i})\sgn(L_{s+1,i+2^s}), \nonumber\\
&L_{s,i+2^s} \!=\! (1-2\hat{v}_{s,i})L_{s+1,i}+L_{s+1,i+2^s},
\label{equ:SCPE_L}
\end{align}
where 
$L_{s,i}$ and $\hat{v}_{s,i}$ are the LLR value and the hard bit estimation at the $s$-th stage, $0 \leq s \leq n$, and the $i$-th bit, $0 \leq i \leq N-1$, respectively. The hard bit values of the PE are computed as
\begin{equation}
\label{equ:SCPE_v}
\begin{aligned}
&\hat{v}_{s+1,i} = \hat{v}_{s,i} \oplus \hat{v}_{s,i+2^s},\\
&\hat{v}_{s+1,i+2^s} = \hat{v}_{s,i+2^s},
\end{aligned}
\end{equation}
where $\oplus$ denotes the logical XOR operation.

The LLR values at the $n$-th stage are initialized to $\bm{L}_n$. In SC decoding, the hard bit estimations at the $0$-th stage are calculated as
\begin{equation}
\label{equ:SC:HardDecision}
\hat{u}_{i} = \hat{v}_{0,i}=
\begin{cases}
0 & \text{if } i \in \mathcal{A}^c,\\
\frac{1 - \sgn(L_{0,i})}{2} & \text{otherwise.}
\end{cases}
\end{equation}
In SCL decoding, at the $0$-th stage, each information bit is estimated as either $0$ or $1$ and at each decoding step, only $M$ most likely candidate paths are allowed to survive. After the last bit is estimated in SCL decoding, the path with the highest reliability metric is selected as the decoding result. If a CRC of length $c$ is used to help SCL decoding, after the last bit is estimated, the path that passes the CRC verification is selected as the decoding result.

\subsection{SCF and DSCF Decoding}
\label{sec:polar:DSCF}

SCF decoding is used to decode a polar code that is concatenated with a CRC of length $c$ for verification. It starts by performing SC decoding and if the CRC verification fails after the initial SC decoding, it flips the bit estimation of an information bit which has the smallest absolute LLR value \cite{SCF}. However, this simple bit-flipping metric prevents SCF decoding to obtain a satisfactory error-correction performance \cite{DSCF}.

To determine the bit-flipping position, DSCF decoding estimates the probability $P^*_{i_\omega}$ of the $i_\omega$-th bit ($i_\omega \in \mathcal{A}$) being the first-order error bit after the initial SC decoding attempt as
\begin{equation}
\label{equ:DSCF:P_flip}
P^*_{i_\omega}=(1 - p^*_{i_\omega}) \times \prod_{\substack{{\forall i \in \mathcal{A} \setminus{i_\omega}}\\
i < i_\omega}} p^*_i,
\end{equation}
where $p^*_i$ is defined as
\begin{equation}
\label{equ:DSCF:p_star}
p^*_i = \text{Pr}(\hat{u}_i=u_i| \bm{y},  \bm{\hat{u}}_0^{i-1} = \bm{u}_0^{i-1}),
\end{equation}
with $\bm{\hat{u}}_0^{i-1} = \{\hat{u}_{0}, \hat{u}_{1}, \dots, \hat{u}_{i-1}\}$, $\bm{u}_0^{i-1} = \{u_0,u_1,\dots,u_{i-1}\}$. Therefore, the bit-flipping position $i^*_\omega$ that maximizes the probability of $\bm{\hat{u}}$ being correctly decoded after the second SC decoding attempt can be calculated as
\begin{equation}
\label{equ:DSCF:flip}
i^*_\omega = \argmax_{\forall i_\omega \in \mathcal{A}} P^*_{i_\omega}.
\end{equation}

Note that $p^*_i$ cannot be obtained during the course of decoding since the message word $\bm{u}$ is unknown to the decoder \cite{DSCF}. Therefore, DSCF approximates $p^*_i$ as
\begin{equation}
\label{equ:DSCF:p_i}
\begin{split}
p^*_i
&\approx \max \left(\text{Pr}(\hat{u}_i=0| \bm{y},\bm{\hat{u}}_0^{i-1}),\text{Pr}(\hat{u}_i=1| \bm{y}, \bm{\hat{u}}_0^{i-1})\right) \\
&= \frac{1}{1+\exp\left(-|L_{0,i}|\right)}.\\
\end{split}
\end{equation}
It was observed in \cite{DSCF} that the approximation in (\ref{equ:DSCF:p_i}) does not result in a desirable error-correction performance. Therefore, a perturbation parameter $\alpha \in \mathbb{R}^+$ is introduced to obtain a better estimation of $p^*_i$ as
\begin{equation}
\label{equ:DSCF:p_star_estimate}
p^*_i \approx \frac{1}{1+\exp\left(-\alpha|L_{0,i}|\right)}\text{.}
\end{equation}

To enable numerically stable computations for a hardware implementation, the bit-flipping metric is defined as \cite{DSCF}
\begin{equation}
\label{equ:DSCF:metric}
\begin{split}
Q_{\text{DSCF}}(i_\omega) & = -\frac{1}{\alpha}\ln(P^*_{i_\omega}) \\
& = |L_{0,i_\omega}| + \sum_{\substack{{\forall i \in \mathcal{A}}\\ i \leq i_\omega}} \frac{1}{\alpha}\ln{\left(1+\exp\left(-\alpha|L_{0,i}|\right)\right)}.
\end{split}
\end{equation}
Consequently, the most probable bit-flipping position $i^*_\omega$ under DSCF decoding can be found as
\begin{equation}
\label{equ:DSCF:sel}
i^*_\omega = \argmin_{\forall i_\omega \in \mathcal{A}} Q_{\text{DSCF}}(i_\omega).
\vspace{1pt}
\end{equation}

In this paper, all the presented decoders only target the first-order error bit. However, the bit-flipping selection schemes presented in this paper can be directly extended to cover the cases of high-order error bits \cite{DSCF,Doan_SiPS19}.

\section{Deep-Learning-Aided Successive-Cancellation Decoding}
\label{sec:DL-SCL}

In this section, a general bit-flipping algorithm for SCL decoding of polar codes is proposed, with the special case of the bit-flipping algorithm for SC decoding when the list size is $1$. Moreover, a new bit-flipping metric is derived that directly utilizes the correlations of the information bits in terms of the likelihood that an information bit is correctly decoded. A training framework is then introduced as the optimization scheme to design the decoder's parameters followed by the evaluation of the proposed scheme.

\subsection{A Bit-Flipping Algorithm for SCL decoding}
\label{sec:DL-SCL:algorithm}

Consider a failure in the SCL decoding with list size $M$ as the SCL decoding attempt in which all the $M$ decoding paths fail the CRC verification. Let $\bm{\hat{u}}[m]$, $0\leq m < M$, be the $m$-th candidate path after the first SCL decoding, $\bm{\hat{u}}[0]$ be the best path after the first SCL decoding attempt, i.e. the path with the smallest path metric \cite{Alexios_LLR_SCLD}, and let $i^*_\omega$ be the estimated first erroneous bit of $\bm{\hat{u}}[0]$. In the proposed scheme, a secondary SCL decoding attempt is performed by keeping only $\bm{\hat{u}}[0]$ and fixing all the information bits before the $i^*_\omega$-th bit. This is because all the estimated information bits before $i^*_\omega$-th are believed to be correct, and the $i^*_\omega$-th bit is flipped to correct the first error bit of $\bm{\hat{u}}[0]$.

The information bits for the second SCL decoding attempt up to the $i^*_\omega$-th bit are fixed as
\begin{equation}
\hat{u}[m]_i=
\begin{cases}
\hat{u}[0]_i &\text{ if } i \in \mathcal{A}, i < i^*_\omega,\\
1-\hat{u}[0]_i &\text{ if } i \in \mathcal{A}, i = i^*_\omega,
\end{cases}
\end{equation}
for $0 \leq m < M$. After the $i^*_\omega$-th information bit, the conventional SCL decoding procedure is performed by estimating each information bit $i>i^*_\omega, i\in \mathcal{A}$ as both $0$ and $1$ and by keeping the best $M$ paths at each decoding step. The path metrics of all the decoding paths are then given as \cite{Alexios_LLR_SCLD}
\begin{equation}
\PM[m]_i=\PM[m]_{i-1}+\Delta,
\end{equation}
where $0 \leq i < N$, $\PM[m]_{-1}=0$, and $\Delta \ge 0$ is the path metric penalty at the $i$-th bit that is calculated as
\begin{equation}
\Delta =
\begin{cases}
\frac{|L[m]_{0,i}|(1-\sgn(L[m]_{0,i}))}{2} & \text{ if } i \in \mathcal{A}^c,\\
\frac{|L[m]_{0,i}|(1-(1-2\hat{u}[m]_i)\sgn(L[m]_{0,i}))}{2} & \text{ otherwise,}
\end{cases}
\end{equation}
where $L[m]_{0,i}$ is the LLR value of the $i$-th bit at the $m$-th path.

Note that the bit-flipping metric of DSCF can be used to estimate $i^*_\omega$. However, this approach requires costly logarithmic and exponential functions, hence they are not attractive for an efficient hardware implementation. In the next subsection, a novel bit-flipping metric is proposed that only requires multiplication and addition operations.

\subsection{The Proposed Bit-Flipping Metric}

Unlike the DSCF decoder which relies on the estimation of the probability $p^*_i, \forall i\in \mathcal{A}$, for the bit-flipping metric computation, a method is proposed to directly estimate the following likelihood ratio
\begin{equation}
l^*_{i_{\omega}} =\max\left\{
\frac{\text{Pr}(\hat{u}[0]_{i_{\omega}}=0|\bm{y},\bm{u})}{\text{Pr}(\hat{u}[0]_{i_{\omega}}=1|\bm{y},\bm{u})},
\frac{\text{Pr}(\hat{u}[0]_{i_\omega}=1|\bm{y},\bm{u})}{\text{Pr}(\hat{u}[0]_{i_\omega}=0|\bm{y},\bm{u})}
\right\}.
\end{equation}
The value $l^*_{i_\omega}$ indicates how likely the estimated message bit $\hat{u}[0]_{i_\omega}$ is correctly decoded given the received signal $\bm{y}$ and the message word $\bm{u}$. The bit index $i^*_\omega$ which is most likely to be the first-order erroneous bit is then obtained as
\begin{equation}
\label{equ:DL-SCL:ll_sel}
i^*_\omega = \argmin_{\forall i_\omega \in \mathcal{A}} l^*_{i_\omega}.
\end{equation}
Similar to $p^*_i$, the value of $l^*_{i_\omega}$ is not available during the decoding process as $\bm{u}$ remains unknown to the decoder. Therefore, the following hypothesis is proposed for the estimation of $l^*_{i_\omega}$:
\begin{equation}
\label{equ:DL-SCL:hypothesis}
l^*_{i_\omega} \approx \prod_{\forall i \in \mathcal{A}} l_i^{\beta_{i_\omega,i}},
\end{equation}
where
\begin{equation}
\label{equ:DL-SCL:l_i}
\begin{split}
l_i & = \max\left\{
\frac{\text{Pr}(\hat{u}_i=0|\bm{y},\bm{\hat{u}}^{i-1}_0)}{\text{Pr}(\hat{u}_i=1|\bm{y},\bm{\hat{u}}^{i-1}_0)},
\frac{\text{Pr}(\hat{u}_i=1|\bm{y},\bm{\hat{u}}^{i-1}_0)}{\text{Pr}(\hat{u}_i=0|\bm{y},\bm{\hat{u}}^{i-1}_0)}
\right\} \\
& = \exp{|L_{0,i}|},
\end{split}
\end{equation}
and $\beta_{i_\omega,i}\in \mathbb{R}$ are perturbation parameters such that ${\beta_{i_\omega,i} = \beta_{i,i_\omega}}$ and $\beta_{i_\omega,i_\omega}=1$, for $i \in \mathcal{A}$ and $i_\omega \in \mathcal{A}$.

To enable numerically stable computations, the bit-flipping metric of the proposed decoder can be obtained by transforming the likelihood ratio $l^*_{i_\omega}$ to the LLR domain as
\begin{equation}
\label{equ:DL-SCL:metric}
\begin{split}
Q_{\text{DL-SCL}}(i_\omega) & = \ln(l^*_{i_\omega}) \\
& \approx \ln \left(\prod_{\forall i \in \mathcal{A}} \exp{\left(\beta_{i_\omega,i}|L_{0,i}|\right)}\right) \\
& = \sum_{\forall i \in \mathcal{A}}\beta_{i_\omega,i} |L_{0,i}|.
\end{split}
\end{equation}
The most probable bit-flipping index $i^*_\omega$ can then be selected as
\begin{equation}
\label{equ:DL-SCL:llr_sel}
i^*_\omega = \argmin_{\forall i_\omega \in \mathcal{A}} Q_{\text{DL-SCL}}(i_\omega).
\end{equation}


Note that the bit-flipping metric computation in (\ref{equ:DL-SCL:metric}) can be represented in the matrix form as
\begin{equation}
\label{equ:DL-SCL:metric_matrix}
\bm{Q}_{\text{DL-SCL}} = |\bm{L}_0| \cdot \bm{\beta},
\end{equation}
where $\bm{Q}_{\text{DL-SCL}}$ and $\bm{L}_0$ are row vectors of size ${1\times(K+c)}$, and $\bm{\beta}$ is a matrix of size $(K+c)\times(K+c)$. Equivalently, $i^*_\omega$ is the index of the element in $\bm{Q}_{\text{DL-SCL}}$ that has the smallest value.

With $\beta_{i_\omega,i_\omega}=1$ and $\beta_{i_\omega,i}=\beta_{i,i_\omega}$, $0 \leq i, i_\omega < K+c$, $\bm{\beta}$ can be seen as a correlation matrix that captures the inherent relations of the absolute LLR values of all the information bits under SCL decoding. For the sake of simplicity, since only the LLR values of information bits are considered, all of the bit indices used in the rest of this paper are referred to as information bit indices, and therefore have the values in the range of $[0,K+c-1]$.

\subsection{Parameter Optimization}

Note that $\beta_{i_\omega,i_\omega}$ is fixed to 1 and is not trainable for all $0 \le i_\omega < K+c$. On the other hand, other elements of the matrix $\bm{\beta}$ are trainable with a condition that ${\beta_{i_\omega,i} = \beta_{i,i_\omega}}$, ${0 \le i,i_\omega < K+c}$. In the proposed DL-SCL decoding, the number of trainable parameters of the matrix $\bm{\beta}$ is $\frac{(K+c)(K+c-1)}{2}$, which is too large to efficiently apply heuristic methods such as Monte Carlo simulation for parameter optimization. Therefore, the optimization of $\bm{\beta}$ is considered as a learning problem and DL techniques are exploited to optimize $\bm{\beta}$. The bit-flipping metric $\bm{Q}_\text{DL-SCL}$ of the proposed decoder does not depend on the values of the message word $\bm{u}$. Thus, all-zero codewords can be used during the training phase. This symmetric property is particularly useful for DL-based decoders of linear block codes, as it simplifies the training process \cite{Doan_SiPS19,Doan_ICC19,Nachmani_STSP}.

Let $\bm{\hat{T}}$ be the estimated bit-flipping vector of the information bits with $-1$ indicating a bit-flip and $+1$ indicating no bit-flip. From (\ref{equ:DL-SCL:llr_sel}), the elements of the vector $\bm{\hat{T}}$ are defined as
\begin{equation}
\label{equ:DL-SCL:T_hat}
\hat{T}_i = 
\begin{cases}
-1 & \text{if } i=i^*_\omega,\\
+1 & \text{if } i \neq i^*_\omega,
\end{cases}
\end{equation}
for $0\leq i < K+c$. In this paper, stochastic-gradient-descent (SGD) based techniques are used to update the values of $\bm{\beta}$ during training, thus the computation of $\bm{\hat{T}}$ is modified to enable back-propagation during training \cite{Doan_SiPS19}. Otherwise, learning is not feasible as the derivative of (\ref{equ:DL-SCL:T_hat}) with respect to $\bm{Q}_\text{DL-SCL}[i]$, i.e., the $i$-th element of $\bm{Q}_\text{DL-SCL}$, is always 0.

Let the soft estimation of $\hat{T}_i$ be
\begin{equation}
\tilde{T}_i = \tanh(\bm{Q}_\text{DL-SCL}[i]-\tau),
\end{equation}
where $\tau=\frac{\tau_0+\tau_1}{2}$, and $\tau_0$ and $\tau_1$ are the smallest and the second smallest values of $\bm{Q}_\text{DL-SCL}$, respectively. The objective loss function is then defined as
\begin{equation}
\label{equ:DL-SCL:train_loss}
\begin{split}
\text{Loss} & = \frac{1}{K+c}\sum_{i=0}^{K+c-1} \mathcal{L}\left(\frac{1-\tilde{T}_i}{2},\frac{1-T_i}{2}\right)\\
& + \lambda \sum_{i_\omega=0}^{K+c-1}\sum_{i=i_\omega+1}^{K+c-1}(\beta_{i_\omega,i})^2,
\end{split}
\end{equation}
where ${\mathcal{L}(a,b)=-b\log(a)-(1-b)\log(1-a)}$ is the binary cross-entropy function, and $\lambda$ is the scaling factor of the L2 regularization \cite{DeepLearning}.

In this paper, PyTorch \cite{PyTorch} is used as the DL framework. Training is done using RMSprop optimizer \cite{hinton2012neural} with a mini-batch size of $128$ and a learning rate of $10^{-4}$. The training set consists of $2^{18}$ samples of the received channel signals, which are not correctly decoded after the first SCL decoding attempt, and the data is collected at ${E_b/N_0=5}$~dB. The L2 regularization hyperparameter $\lambda$ is set to $0.25$. The initial values of the non-diagonal elements of $\bm{\beta}$ are drawn from an i.i.d distribution within the range of $[-0.2,0.2]$ before training takes place. The matrix $\bm{\beta}$ is trained for list sizes $M\in \{1,2,4,8\}$\footnote{Optimized $\bm{\beta}$ matrices are available at https://github.com/nghiadt05/DLSCL-CorMatrices}.
 
\subsection{Evaluation}

\begin{figure}[t]
	\centering
	\begin{tikzpicture}[spy using outlines = {rectangle, magnification=2.0, connect spies}]

\pgfplotsset{	
	label style = {font=\fontsize{7pt}{7}\selectfont},
	tick label style = {font=\fontsize{7pt}{7}\selectfont}
}

\begin{axis}[
scale = 1,
ymode=log,
xlabel={$E_b/N_0$ [\text{dB}]}, 
ylabel={FER}, 
xtick={4,4.5,5,5.5,6,6.5},
ytick={1e-5,1e-4,1e-3,1e-2,1e-1},
grid=both,
ymajorgrids=true,
xmajorgrids=true,
grid style=dashdotted,
width=1\columnwidth, 
height=6.5cm,
thick,
mark size=2,
legend cell align={left},
legend style={
	at={(0,1e-5)},
	anchor=south west,
	column sep= 2mm,
	font=\fontsize{6pt}{7.2}\selectfont,
},
  legend to name=perf-legend-DLSCL,
  legend columns=4,
]

\addplot[
color=red,
solid,
mark=*,
every mark/.append style={solid,fill=red},
thick,
mark size=2,
]
table {
	4	0.03564453
	4.25	0.01953125
	4.5	0.0095825195
	4.75	0.0050354004
	5	0.0020811986
	5.25	0.0008802817
	5.5	0.00035386675
	5.75	0.00011312217
	6	4.19E-05
};
\addlegendentry{DL-SCL1}

\addplot[
color=red,
solid,
mark=triangle*,
every mark/.append style={solid,fill=red},
thick,
mark size=2,
]
table {
	4	0.01473999
	4.25	0.0066833496
	4.5	0.003092448
	4.75	0.0012626263
	5	0.00040849674
	5.25	0.00013455328
	5.5	5.11E-05
};
\addlegendentry{DL-SCL2}

\addplot[
color=red,
solid,
mark=square*,
every mark/.append style={solid,fill=red},
thick,
mark size=2,
]
table {
	4	0.0072631836
	4.25	0.0029761905
	4.5	0.0011261262
	4.75	0.00036873156
	5	0.00011170688
	5.25	3.83E-05
};
\addlegendentry{DL-SCL4}

\addplot[
color=red,
solid,
mark=diamond*,
every mark/.append style={solid,fill=red},
thick,
mark size=2,
]
table {
	4	0.0034722222
	4.25	0.001344086
	4.5	0.0004241635
	4.75	0.000117370895
	5	4.27E-05
};
\addlegendentry{DL-SCL8}

\addplot[
color=blue,
dotted,
every mark/.append style={solid},
mark=o,
thick,
mark size=2,
]
table {
	4	0.035491943
	4.25	0.019439697
	4.5	0.009552002
	4.75	0.005004883
	5	0.0020651894
	5.25	0.00087412586
	5.5	0.0003511236
	5.75	0.000112410075
	6	4.07E-05
};
\addlegendentry{SCLF1 \cite{DSCF}}

\addplot[
color=blue,
dotted,
every mark/.append style={solid},
mark=triangle,
thick,
mark size=2,
]
table {
	4	0.014556885
	4.25	0.0065612793
	4.5	0.003045945
	4.75	0.0012350644
	5	0.00039184952
	5.25	0.0001320509
	5.5	5.05E-05
};
\addlegendentry{SCLF2 \cite{DSCF}}

\addplot[
color=blue,
dotted,
every mark/.append style={solid},
mark=square,
thick,
mark size=2,
]
table {
	4	0.0071411133
	4.25	0.0029296875
	4.5	0.001067598
	4.75	0.00036231885
	5	0.00010407993
	5.25	3.38E-05
};
\addlegendentry{SCLF4 \cite{DSCF}}

\addplot[
color=blue,
dotted,
every mark/.append style={solid},
mark=diamond,
thick,
mark size=2,
]
table {
	4	0.0033151726
	4.25	0.0012376237
	4.5	0.0004071661
	4.75	0.00010945709
	5	3.83E-05
};
\addlegendentry{SCLF8 \cite{DSCF}}

\addplot[
color=black,
solid,
mark=o,
thick,
mark size=2,
]
table {
	4	0.14987183
	4.25	0.10006714
	4.5	0.06561279
	4.75	0.0395813
	5	0.024017334
	5.25	0.014190674
	5.5	0.007843018
	5.75	0.0038470645
	6	0.0022904829
	6.25	0.0011422294
	6.5	0.00056053814
};
\addlegendentry{SCL1 \cite{Alexios_LLR_SCLD}}

\addplot[
color=black,
solid,
mark=triangle,
thick,
mark size=2,
]
table {
	4	0.05682373
	4.25	0.030517578
	4.5	0.017089844
	4.75	0.0076293945
	5	0.0035536024
	5.25	0.0016575863
	5.5	0.0007296157
	5.75	0.00025201612
	6	9.04E-05
};
\addlegendentry{SCL2 \cite{Alexios_LLR_SCLD}}

\addplot[
color=black,
solid,
mark=square,
thick,
mark size=2,
]
table {
	4	0.026184082
	4.25	0.013153076
	4.5	0.006164551
	4.75	0.002395342
	5	0.00086206896
	5.25	0.00029904305
	5.5	0.00010841284
};
\addlegendentry{SCL4 \cite{Alexios_LLR_SCLD}}

\addplot[
color=black,
solid,
mark=diamond,
thick,
mark size=2,
]
table {
	4	0.013214111
	4.25	0.005584717
	4.5	0.002227248
	4.75	0.0007267442
	5	0.00023496241
	5.25	6.31E-05
};
\addlegendentry{SCL8 \cite{Alexios_LLR_SCLD}}

\end{axis}
\end{tikzpicture}
	\ref{perf-legend-DLSCL}
	\caption{FER comparison of various decoders for $\mathcal{P}(128,64)$ concatenated with a $24$-bit CRC.}
	\label{fig:FER:all}
\end{figure}
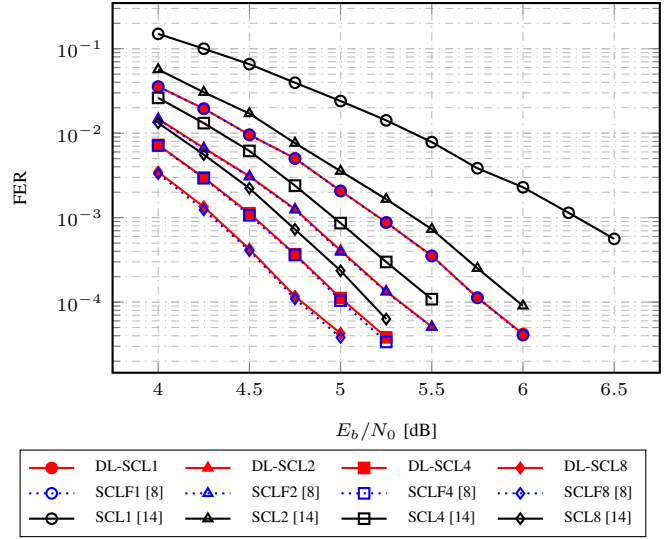


In this section, the performance of the proposed DL-SCL decoder in terms of frame-error-rate (FER) and computational complexity is examined. The polar code $\mathcal{P}(128,64)$ concatenated with a $24$-bit CRC is considered. The selected polar code and the CRC polynomial are used in the eMBB control channel of the 5G standard \cite{3gpp_report}.

Fig.~\ref{fig:FER:all} compares the FER performance of various decoders for $\mathcal{P}(128,64)$. In this figure, DL-SCL$M$ denotes the proposed DL-SCL decoding algorithm with list size $M=\{1,2,4,8\}$, and the bit-flipping SCL decoder with the bit-flipping metric proposed in \cite{DSCF} is denoted as SCLF$M$. In addition, the original SCL decoding in \cite{Alexios_LLR_SCLD} is also considered for the comparison. For all the bit-flipping SCL decoders, $8$ additional decoding attempts are considered for the secondary SCL decoding. As observed from Fig.~\ref{fig:FER:all}, the proposed bit-flipping metric in the proposed DL-SCL decoders results in almost no FER performance loss compared to that of the SCLF decoders.

\begin{figure}[t]
	\centering
	\includegraphics[width=.8\linewidth]{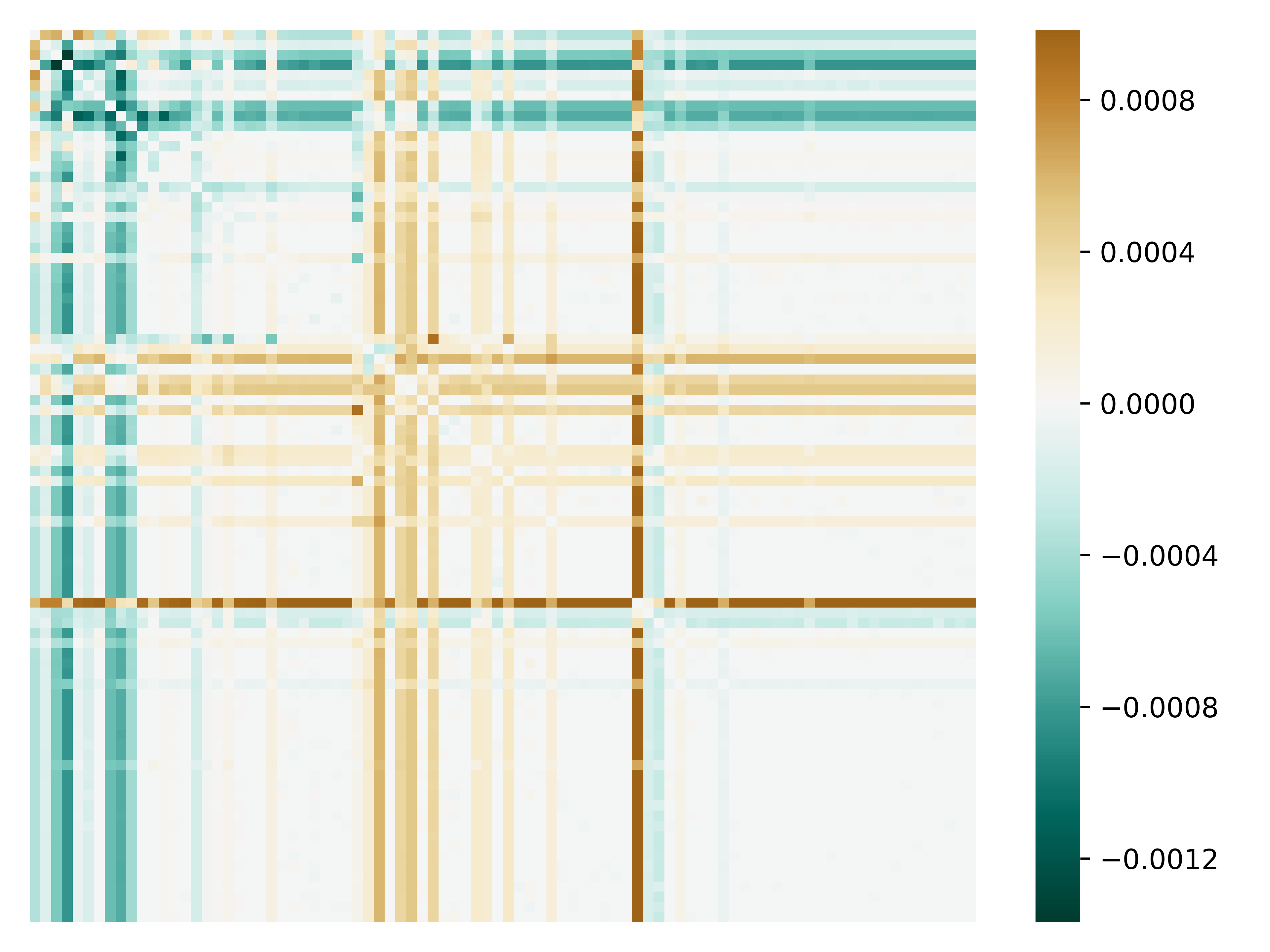}
	\caption{Visualization of $\bm{\beta} - \bm{I}$ for the DL-SCL$8$ decoder.}
	\label{fig:Gamma}
\end{figure}

Fig.~\ref{fig:Gamma} visualizes the values of the elements in matrix $\bm{\beta} - \bm{I}$ in the form of a heat map\footnote{Matrix $\bm{\beta} - \bm{I}$ is shown to exclude the diagonal elements of $\bm{\beta}$ which have a value of $1$.}, where $\bm{I}$ is the identity matrix with the same size as $\bm{\beta}$. It can be seen that $\bm{\beta} - \bm{I}$ (and thus $\bm{\beta}$) is a sparse matrix with many of its elements having a value close to $0$. This observation is exploited to reduce the computational complexity of computing the proposed bit-flipping metric, which in turn reduces the computational complexity of the proposed DL-SCL decoding algorithm.

Table~\ref{tab:complexity_count} reports the computational complexity of the proposed bit-flipping metric of the DL-SCL decoder in comparison with that of the SCLF decoder in terms of the number of different operations performed. Other than the bit-flipping metric, the proposed DL-SCL decoder and the SCLF decoder are identical. In this table, all the elements of $\bm{\beta}$ which have a value in the range $[-10^{-4},10^{-4}]$ are set to $0$, thus removing the need to perform additions or multiplications over those elements, without tangibly degrading the error-correction performance. It can be seen that the bit-flipping metric computation in the proposed DL-SCL decoders require up to $66\%$ fewer multiplications and up to $36\%$ fewer additions in comparison with that of the SCLF decoder. Moreover, unlike SCLF decoder, the proposed bit-flipping metric in the DL-SCL decoders does not require the computation of any transcendental functions.



\begin{table}[t]
	\caption{Computational complexity of the bit-flipping metric for $\mathcal{P}(128,64)$ in terms of the number of operations performed}
	\def\arraystretch{1.25}
	\centering
	\begin{tabular}{l c c c}
		\toprule
		Decoders & $\times$ & $+$ & $\ln$/$\exp$ \\
		\midrule
		SCLF$M$ & $7832$ & $4004$ & $7832$ \\
		DL-SCL$1$ & $2652$ & $2564$ & $0$ \\
		DL-SCL$2$ & $3116$ & $3028$ & $0$ \\
		DL-SCL$4$ & $3238$ & $3150$ & $0$ \\
		DL-SCL$8$ & $3176$ & $3088$ & $0$ \\
		\bottomrule
	\end{tabular}
	\label{tab:complexity_count}
\end{table}

\section{Conclusion}
\label{sec:summary}

In this paper, a deep-learning-aided successive-cancellation list (DL-SCL) decoding algorithm for polar codes is introduced. The proposed decoder improves the performance of successive-cancellation list (SCL) decoding by running additional SCL decoding attempts using a novel bit-flipping scheme. The bit-flipping metric of the proposed decoder is obtained by exploiting the inherent relations between the information bits. These relations are  expressed in the form of a trainable correlation matrix, which is optimized using deep-learning (DL) techniques. Performance results on a polar code of length $128$ and rate $1/2$ show that the proposed bit-flipping metric in the proposed DL-SCL decoder requires up to $66\%$ fewer multiplications and up to $36\%$ fewer additions in comparison with the state of the art, while providing almost the same error-correction performance.

\section*{Acknowledgment}
S. A. Hashemi is supported by a Postdoctoral Fellowship from the Natural Sciences and Engineering Research Council of Canada (NSERC).


\end{document}